\documentclass[10pt, conference, letterpaper]{IEEEtran}
\usepackage[letterpaper, left = 0.68in, right =0.68in, top = 0.7in, bottom = 1in]{geometry}
\usepackage[noadjust]{cite}
\usepackage{amsmath,amssymb,amsfonts}
\usepackage{graphicx}
\usepackage{textcomp}
\usepackage{mathtools}
\usepackage{adjustbox}
\usepackage{bm}
\usepackage{tikz}
\usepackage{hhline}
\usetikzlibrary{arrows, fit, matrix, positioning, shapes, backgrounds}
\usepackage{xcolor}
\def\BibTeX{{\rm B\kern-.05em{\sc i\kern-.025em b}\kern-.08em
    T\kern-.1667em\lower.7ex\hbox{E}\kern-.125emX}}

\usepackage{supertabular}    
\usepackage{array} 
\usepackage[linesnumbered,ruled]{algorithm2e}
\usepackage{algorithmic}
\usepackage{todonotes}
\usepackage{numprint}     
\usepackage{tikz}
\usepackage{makecell}   
\usepackage{xcolor,colortbl}
\usepackage{enumitem}            
\usepackage{longtable} 
\usepackage{wrapfig}
\usepackage{subcaption}
\usepackage{pgfplots}
\usepgfplotslibrary{groupplots}
\usepackage{booktabs}
\usepackage{graphics}
\usepackage{makecell}
\usepackage[normalem]{ulem}
\usepackage{tikz}
\usepackage{pgfplots}
\usepackage{pgfplotstable}
\usetikzlibrary{external}
\usepackage{amsthm}
\usepackage{amsmath, amssymb}

\pgfplotsset{vasymptote/.style={
    before end axis/.append code={
        \draw[densely dashed, line width=1.5pt] ({rel axis cs:0,0} -| {axis cs:#1,0})
        -- ({rel axis cs:0,1} -| {axis cs:#1,0});
    }
}}

\DeclareMathOperator*{\argmax}{arg\,max}

\usepackage[utf8]{inputenc}
\usepackage[acronym]{glossaries}

\newcommand{\egc}{e.\,g., }
\newcommand{\iec}{i.\,e., }

\newcommand{\wrt}{w.\,r.\,t.\ }

\newcolumntype{?}{!{\vrule width 1pt}}
\definecolor{mittelblau}{RGB}{0, 126, 198}
\definecolor{violettblau}{cmyk}{0.9, 0.6, 0, 0}
\definecolor{rot}{RGB}{238, 28 35}
\definecolor{apfelgruen}{RGB}{140, 198, 62}
\definecolor{gelb}{RGB}{1, 221, 0}
\definecolor{orange}{RGB}{244, 111, 33}
\definecolor{pink}{RGB}{237, 0, 140}
\definecolor{lila}{RGB}{128, 10, 145}
\definecolor{hellgrau}{RGB}{224, 224, 224}
\definecolor{mittelgrau}{RGB}{128, 128, 128}
\definecolor{dunkelgrau}{RGB}{80,80,80}
\definecolor{anthrazit}{RGB}{19, 31, 31}
\definecolor{darkgreen}{RGB}{0.125,0.5,0.169}

\begin{document}

\title{Probabilistic 5G Indoor Positioning Proof of Concept with Outlier Rejection}

\author{
    \IEEEauthorblockN{
        Marcus Henninger\IEEEauthorrefmark{1}\IEEEauthorrefmark{2},
        Traian E. Abrudan\IEEEauthorrefmark{1},
        Silvio Mandelli\IEEEauthorrefmark{1},
        Maximilian Arnold\IEEEauthorrefmark{1},
        Stephan Saur\IEEEauthorrefmark{1}, \\
        Veli-Matti Kolmonen\IEEEauthorrefmark{1},
        Siegfried Klein\IEEEauthorrefmark{1}, 
        Thomas Schlitter\IEEEauthorrefmark{1},
        and Stephan ten Brink\IEEEauthorrefmark{2}
        }

	\IEEEauthorblockA{
	\IEEEauthorrefmark{1}Nokia Bell Labs\\
	\IEEEauthorrefmark{2}Institute of Telecommunications, University of Stuttgart, 70659 Stuttgart, Germany \\
	E-mail: marcus.henninger@nokia.com, \{firstname.lastname\}@nokia-bell-labs.com}}

\maketitle

\newacronym{1D}{1D}{one-dimensional}
\newacronym{2D}{2D}{two-dimensional}
\newacronym{3D}{3D}{three-dimensional}
\newacronym{5G}{5G}{5th generation}
\newacronym[plural=AoAs,firstplural = angles of arrival (AoAs)]{aoa}{AoA}{angle of arrival}
\newacronym{awgn}{AWGN}{additive white Gaussian noise}
\newacronym{cdf}{CDF}{cumulative distribution function}
\newacronym{csi}{CSI}{channel state information}
\newacronym{irls}{IRLS}{iteratively reweighted least squares}
\newacronym{isac}{ISAC}{integrated sensing and communication}
\newacronym{ll}{LL}{log-likelihood}
\newacronym{ml}{ML}{maximum likelihood}
\newacronym{music}{MUSIC}{MUltiple SIgnal Classification}
\newacronym{nlos}{NLOS}{non-line of sight}
\newacronym{nr}{NR}{New Radio}
\newacronym{ofdm}{OFDM}{orthogonal frequency-division multiplexing}
\newacronym{rmse}{RMSE}{root-mean-square error}
\newacronym{rss}{RSS}{received signal strength}
\newacronym{snr}{SNR}{signal-to-noise ratio}
\newacronym{tdoa}{TDoA}{time difference of arrival}
\newacronym{toa}{ToA}{time of arrival}
\newacronym{ula}{ULA}{uniform linear array}
\newacronym{ue}{UE}{user equipment}
\newacronym{vmf}{VMF}{von Mises-Fisher}
\newacronym{wls}{WLS}{weighted least squares}

\begin{abstract}
The continuously increasing bandwidth and antenna aperture available in wireless networks laid the foundation for developing competitive positioning solutions relying on communications standards and hardware. However, poor propagation conditions such as \gls{nlos} and rich multipath still pose many challenges due to outlier measurements that significantly degrade the positioning performance.

In this work, we introduce an iterative positioning method that reweights the \gls{toa} and \gls{aoa} measurements originating from multiple locators in order to efficiently remove outliers. In contrast to existing approaches that typically rely on a single locator to set the time reference for the \gls{tdoa} measurements corresponding to the remaining locators, and whose measurements may be unreliable, the proposed iterative approach does not rely on a reference locator only. The resulting robust position estimate is then used to initialize a computationally efficient gradient search to perform maximum likelihood position estimation. 

Our proposal is validated with an experimental setup at 3.75~GHz with 5G numerology in an indoor factory scenario, achieving an error of less than 50~cm in 95\% of the measurements. To the best of our knowledge, this paper describes the first proof of concept for 5G-based joint \gls{toa} and \gls{aoa} localization. 
\end{abstract}

\vspace{0.25cm}


\glsresetall
\makeatletter{\renewcommand*{\@makefnmark}{}
\footnotetext{\copyright\;2022 IEEE. Personal use of this material is permitted. Permission from IEEE must be obtained for all other uses, in any current or future media, including reprinting/republishing this material for advertising or promotional purposes, creating new collective works, for resale or redistribution to servers or lists, or reuse of any copyrighted component of this work in other works.}\makeatother}

\section{Introduction}\label{sec:intro}

Positioning using radio signals has been extensively studied in literature, and will gain further importance due to the emergence of \gls{isac}~\cite{viswanathan2020communications}. Location information is a key enabler of many applications, \egc in the areas of security or healthcare~\cite{zafari2019survey}. The possibility of deploying \gls{5G} systems as private networks makes indoor positioning using 5G signals a particularly interesting task and an attractive technology for enterprise customers. Especially in factory scenarios, accurate positioning will be of crucial importance as it paves the way for fully automated factories, \egc by enabling asset tracking or robot navigation~\cite{de2021convergent}.

In addition to an accurate localization performance overall, the prevention of outliers is pivotal for a positioning system. This requirement becomes especially challenging in indoor environments, where multipath components due to \gls{nlos} propagation frequently occur. Those effects can create outliers in the \gls{toa} and \gls{aoa} measurements, that are the basis for many localization systems, leading to severely impacted position estimates. One way of coping with this issue is, instead of employing \gls{toa}-only or \gls{aoa}-only solutions, to jointly use \gls{toa} and \gls{aoa} measurements to increase the robustness of a system, as done \egc in~\cite{eren2011cooperative, jeong2014joint, naseri2018bayesian}. 
More recently,~\cite{geng2021experimental} has investigated the performance of probabilistic joint \gls{toa}/\gls{aoa} positioning in indoor environments. None of these previous works, however, effectively mitigates the impact of multipath propagation, \egc by detecting and rejecting such components beforehand. 

Similarly to \cite{geng2021experimental}, in this work, we are focusing on probabilistic positioning algorithms due to their excellent performance under appropriate probability distributions assumptions. However, such techniques exhibit the drawback of either prohibitive complexity of multi-dimensional grid search or the requirement of a proper initial search point, that enables gradient-based optimization with its light requirements in terms of computational complexity. 
Moreover, the performance of \gls{ml} algorithms can be enhanced by removing outliers, created \egc due to the already mentioned multipath propagation, and by weighting the remaining measurements based on the confidence in them.
In literature, various approaches based on \gls{tdoa} exist that could be employed to determine an initial search point, \egc \cite{rosic2016tdoa, ma2019tdoa, lee2021iterative, amar2010reference}. However, \cite{rosic2016tdoa, ma2019tdoa, lee2021iterative} require the definition of a reference locator, which has the drawback that the reference itself could exhibit an outlier measurement. On the other hand, \cite{amar2010reference} avoids the definition of a reference locator, but can not detect outliers and does not provide the possibility of assigning weights to the measurements.

In this paper, we present an experimental 5G positioning system based on joint \gls{toa}/\gls{aoa} probabilistic positioning with robust initialization that overcomes the aforementioned drawbacks. 
Through an \gls{irls} technique, our proposal can reject outliers, mitigating the effects of multipath propagation. Moreover, through this iterative approach, we can assign weights to the remaining measurements according to the confidence in them. Our initialization routine effectively acts as a pre-filtering method to improve the performance of the investigated \gls{ml} positioning algorithms. We show that with joint \gls{toa}/\gls{aoa} \gls{ml} positioning in conjunction with initialization, we achieve a \gls{2D} error of at most ca. 50 cm~for 95\% of the measurements conducted in a real-world indoor environment operating at 3.75~GHz carrier frequency with \gls{5G} numerology~\cite{5G_3gpp} and 100~MHz bandwidth. 
To the best of the authors' knowledge, this work demonstrates the first joint \gls{toa}/\gls{aoa} positioning proof of concept operating with \gls{5G} signals.

The rest of the paper is structured as follows: in Section~\ref{sec:system}, we explain how \gls{toa} and \gls{aoa} measurements are taken by locators. Section~\ref{sec:prob_pos} introduces the probabilistic positioning algorithms. The main part of this work is our robust initialization routine that is presented in Section~\ref{sec:init}. In Section~\ref{sec:exp_setup}, we briefly explain the experimental setup that will be used in Section \ref{sec:results} to compare our proposal's performance versus other baselines. Conclusions are drawn in Section~\ref{sec:conclusion}.



\section{\gls{toa} and \gls{aoa} estimation}\label{sec:system}

We consider a positioning system with $K$ locators with indices in the set $\mathcal{K}=\{1,2,\ldots, K\}$, aiming at localizing a \gls{ue}, whose unknown \gls{3D} location is denoted by $\mathbf{x} = [x, y, z]$. The known \gls{3D} position of the $k$-th locator is given by the vector $\mathbf{p}_k = [x_k, y_k, z_k]$. Its \gls{3D} orientation is defined by a $3 \times 3$ matrix $\mathbf{\Omega}_k$, which contains the orthogonal unit vectors corresponding to the locator's orientation \wrt the reference coordinate system.

Each locator independently provides estimates of the \gls{toa}, corresponding to a distance estimate $\hat{d}_k$. Note that to enable \gls{tdoa}-based approaches, we require all locators to be synchronized with each other, but not with the \gls{ue}. Therefore, the \gls{toa} estimates include an unknown transmit time. The estimated azimuth and elevation \glspl{aoa} of the \gls{ue} from each locator are parametrized using a directional statistics approach~\cite{abrudan2016underground}.

The \gls{toa} and \gls{aoa} measurements serve as inputs to our positioning algorithms, which will be described in Sections \ref{sec:prob_pos} and \ref{sec:init}. While \gls{toa} and \gls{aoa} estimation are not the focus of this work due to space reasons, we refer the interested reader 
to~\cite{henninger2021computationally}. The approach of this previous work allows us to obtain estimates of the three parameters of interest by utilizing multi-dimensional \gls{music} in a computationally efficient manner. Managing computational complexity is critical, as a typical 5G signal parametrization comes with a high number of subcarriers \cite{5G_3gpp} and, thus, leads to a large number of samples to be processed.




\section{Probabilistic Positioning Algorithms}\label{sec:prob_pos}
In order to locate the \gls{ue}, we consider the
\gls{toa}-only, \gls{aoa}-only, and joint \gls{toa}/\gls{aoa} \gls{ml} techniques described below. 

\subsection{\gls{toa} Maximum Likelihood Positioning}\label{sec:tdoa_ml}
We model the \gls{toa} estimation error as a zero-mean Gaussian random variable such that the \gls{ll} function of the $K$ locators is given as
\begin{align}
\mathcal{L}_{T}(\mathbf{x}, \tau)
& = \sum_{k \in \mathcal{K}} \ln p(\hat{d}_k\vert\mathbf{x}, \tau) \nonumber \\
& = - \sum_{k \in \mathcal{K}} w_{T, k} \cdot \frac{(\hat{d}_k - \lVert \mathbf{p}_k - \mathbf{x} \rVert - \tau \cdot c)^{2}}{2} 
\label{eq:toa_ll}
\end{align}
where $w_{T, k}$ is the weight of the $k$-th locator, corresponding to the inverse variance of the \gls{toa} estimate $\sigma_k^{-2}$, and the constant term has been omitted. Furthermore, $\tau$ denotes the unknown transmit time and $c$ the speed of light. A position estimate only relying on \gls{toa} measurements is obtained by maximizing the joint \gls{ll} function of the $K$ locators~\cite{perez2016blade}
\begin{align}
(\hat{\mathbf{x}}_{T}, \hat{\tau}_{T})
& = \argmax_{\mathbf{x}, \tau} \mathcal{L}_{T}(\mathbf{x}, \tau)\;.
\label{eq:toa_ll_max}
\end{align}
Note that solving \eqref{eq:toa_ll_max} also yields an estimate of the transmit time $\hat{\tau}_{T}$ as an additional nuisance parameter. 

\subsection{\gls{aoa} Maximum Likelihood Positioning}\label{sec:aoa_ml}
Similarly to the \gls{toa} approach, a position estimate can also be obtained solely based on \gls{aoa} measurements.
We use the \gls{vmf} distribution to model angular uncertainties, as in~\cite{abrudan2016underground}. 
The \gls{ll} function (the constant term has been omitted) of the $K$ locators is written as 
\begin{align}
\mathcal{L}_{\angle}(\mathbf{x})
& = \sum_{k \in \mathcal{K}} \ln \text{VMF}(\mathbf{u}_{k} \vert \hat{\mathbf{u}}_{k}, \kappa_{k}) \nonumber \\
& = 
\sum_{k \in \mathcal{K}} \kappa_{k} \hat{\mathbf{u}}_{k}^{\text{T}} \mathbf{\Omega}_{k}^{\text{T}} 
\frac{\mathbf{x} - \mathbf{p}_{k}}{\lVert \mathbf{x} - \mathbf{p}_{k} \rVert} \; 
\label{eq:vmf_ll}
\end{align}
where $\hat{\mathbf{u}}_{k} \in {\mathbb R}^3$ is a unit vector representing the mean direction in the locator's reference frame $\mathbf{\Omega}_{k}$, obtained by the estimated \glspl{aoa}.
The concentration parameters ${\kappa}_k$ reflect the reliability of the angular measurements and are set according to ${\kappa}_k=\kappa_{\text{max}} w_{\angle,k}$, where $\kappa_{\text{max}}$ is the maximum concentration parameter, which characterizes the minimum spread of the AoA estimates and is hardware specific. The weights $w_{\angle,k}$ are obtained according to reliability information of the angular measurements as described in~\cite{abrudan_2018}.
\\Analogously to \eqref{eq:toa_ll_max}, maximizing \eqref{eq:vmf_ll} yields an \gls{aoa}-based position estimate
\begin{align}
\hat{\mathbf{x}}_{\angle}
& = \argmax_{\mathbf{x}} \mathcal{L}_{\angle}(\mathbf{x}) \; .
\label{eq:vmf_ll_max}
\end{align}

\subsection{Joint \gls{toa}/\gls{aoa} Maximum Likelihood Positioning}\label{sec:joint_ml}

As the third algorithm, we consider \gls{ml} positioning using \gls{toa} and \gls{aoa} in a joint fashion. Assuming \gls{toa} and \gls{aoa} estimates to be independent, \eqref{eq:toa_ll} and \eqref{eq:vmf_ll} are combined to obtain a more robust joint \gls{toa}/\gls{aoa} \gls{ml} position estimate~\cite{abrudan2016underground}
\begin{align}
(\hat{\mathbf{x}}_{\cap}, \hat{\tau}_{\cap})
& = \argmax_{\mathbf{x}, \tau} \mathcal{L}_{\cap}(\mathbf{x}, \tau) \nonumber \\
& = \argmax_{\mathbf{x}, \tau} \big \{ \mathcal{L}_{T}(\mathbf{x}, \tau) +  \mathcal{L}_{\angle}(\mathbf{x})\big \} \; .
\label{eq:joint_ll_fun}
\end{align}



\section{Robust Initialization Routine}\label{sec:init}
Since we rely on gradient-based optimization to maximize the  \gls{ll} functions with affordable complexity, a proper initial search point is necessary to allow convergence to the global maximum. Additionally, the performance of the algorithms can be optimized by rejecting - or weighting less - outlier measurements, and by giving higher confidence to the ones that are deemed more reliable. Towards this end, we introduce a robust initialization routine to maximize the potential of probabilistic positioning. Our routine comprises a novel \gls{irls} \gls{tdoa} algorithm, which is the main contribution of this paper and will be described extensively in what follows. To complement the \gls{tdoa} part, we utilize a prior-art \gls{irls} \gls{aoa} algorithm based on similar principles \cite{abrudan_2018}. These two techniques can be employed separately to initialize the optimization of the \gls{toa} and \gls{aoa} \gls{ml} functions of \eqref{eq:toa_ll_max} and \eqref{eq:vmf_ll_max}, respectively. We then show in Subsection \ref{sec:joint_pos} how we combine their outputs for the initialization of the joint \gls{toa}/\gls{aoa} positioning of \eqref{eq:joint_ll_fun}.

\subsection{\gls{irls} \gls{tdoa} Algorithm}\label{sec:irls_tdoa}

Selecting arbitrary locator $r$ as the reference, $K-1$ distance difference (or \gls{tdoa}, terms to be used interchangeably hereinafter) measurements are constructed as 
\begin{align}
\hat{d}_{k,r} = \hat{d}_{k} - \hat{d}_{r}, \quad k \in \mathcal{K} \backslash \{r\}  \; .
\label{eq:distance_differences}
\end{align}
In order to solve the positioning problem based on these \gls{tdoa} measurements, various prior-art techniques exist in literature. As done \egc in \cite{rosic2016tdoa}, an estimate of the \gls{ue}'s position $\mathbf{x}$ can be computed by solving the linear equation system
\begin{align}
\mathbf{A}_r\mathbf{\Theta}_r = \mathbf{b}_r + \mathbf{n} \; ,
\label{eq:TDoA_LES}
\end{align}
with
\begin{equation}
\mathbf{A}_r =
  \begin{bmatrix}
    (\mathbf{p}_1 - \mathbf{p}_r)^{\text{T}} & \hat{d}_{1, r} \\
    \vdots & \vdots \\
    (\mathbf{p}_{r-1} - \mathbf{p}_r)^{\text{T}} & \hat{d}_{r-1, r} \\
    (\mathbf{p}_{r+1} - \mathbf{p}_r)^{\text{T}} & \hat{d}_{r+1, r} \\
    \vdots & \vdots \\
    (\mathbf{p}_{K} - \mathbf{p}_r)^{\text{T}} & \hat{d}_{K, r} \\
  \end{bmatrix}\; ,
  \label{eq:ls_a}
\end{equation}
\begin{equation}
\mathbf{\Theta}_r = 
\begin{bmatrix} x-x_r & y-y_r & z-z_r & d_r
 \end{bmatrix}^{\text{T}}\; ,
\end{equation}
and
\begin{equation}
\mathbf{b}_r = \frac{1}{2}
  \begin{bmatrix}
    \lVert \mathbf{p}_1 - \mathbf{p}_r \rVert ^{2} - \hat{d}_{1, r}^2 \\
    \vdots \\
    \lVert \mathbf{p}_{r-1} - \mathbf{p}_r \rVert ^{2} - \hat{d}_{r-1, r}^2 \\
    \lVert \mathbf{p}_{r+1} - \mathbf{p}_r \rVert ^{2} - \hat{d}_{r+1, r}^2 \\
    \vdots \\
    \lVert \mathbf{p}_{K} - \mathbf{p}_r \rVert ^{2} - \hat{d}_{K, r}^2 \\
    \end{bmatrix}\; 
  \label{eq:ls_b}
\end{equation}
where $d_r = \lVert \mathbf{x} - \mathbf{p}_r \rVert$, and $\mathbf{n}$ is the noise vector. The closed-form approximate \gls{wls} solution of (\ref{eq:TDoA_LES}) is obtained as
\begin{align}
\hat{\mathbf{\Theta}}_r
 = \big(\mathbf{A}_r^{\text{T}}\mathbf{W}_r\mathbf{A}_r \big) ^{\text{-1}}\mathbf{A}_r^{\text{T}}\mathbf{W}_r\mathbf{b}_r \; 
\label{eq:ls_solution}
\end{align}
with $\mathbf{W}_r = \text{diag}\{w_{T, 1}, \ldots, w_{T, r-1}, w_{T, r+1}, \ldots, w_{T, K}\}$ being the diagonal weighting matrix. The position estimate is then
\begin{align}
\hat{\mathbf{x}}_r
 = \mathbf{p}_r + \hat{\mathbf{\Theta}}_{r, 1:3} \; 
\label{eq:ls_position}
\end{align}
where $\hat{\mathbf{\Theta}}_{r, 1:3}$ is a vector of the first three elements of $\hat{\mathbf{\Theta}}_r$.
The major drawback of the estimator in~\eqref{eq:ls_position} is that it requires the definition of a unique reference locator. If the distance estimate at this locator is an outlier, the position estimate $\hat{\mathbf{x}}_r$ will be severely impacted. We therefore propose an iterative solution that circumvents the need for a single reference locator.
Instead of determining a unique reference locator to compute $\hat{\mathbf{x}}_r$, we use every locator as the reference once to obtain a matrix with $K$ distinct position estimates $\mathbf{\hat{X}} = [\hat{\mathbf{x}}_1, \ldots, \hat{\mathbf{x}}_K]$ utilizing \eqref{eq:TDoA_LES} - \eqref{eq:ls_position}. Then, a weighted average position estimate can be computed as
\begin{align}
\hat{\mathbf{x}}_{\text{WA}}
 = \sum_{k \in \mathcal{K}} \bar{w}_{T, k} \cdot \hat{\mathbf{x}}_k \; 
\label{eq:weighted_position}
\end{align}
where $\bar{w}_{T, k}$ represents the normalized $k$-th locator weight. In every iteration, the weights $w_{T, k}$ are normalized by dividing them by their total sum
\begin{align}
\bar{w}_{T, k} = \frac{w_{T, k}}{\sum_{k \in \mathcal{K}} w_{T, k}}, \quad k \in \mathcal{K}\; .
\label{eq:weight_norm}
\end{align}
For the first iteration, equal weights per locator are used, i.e., $\bar{w}_{T, k} = 1/K, k \in \mathcal{K}$. To detect locators with erroneous measurements, a measure of reliability is required for each of the $K$ position estimates. We therefore define the residual error of each locator's estimate as 
\begin{align}
e_r 
&= \frac{1}{K-1} \sum_{k \in \mathcal{K} \backslash \{r\}} \lvert \hat{d}_{k, r} - (\lVert \hat{\mathbf{x}}_{\text{WA}} - \mathbf{p}_{k} \rVert - \lVert \hat{\mathbf{x}}_{\text{WA}} - \mathbf{p}_r \rVert) \rvert \; 
\label{eq:res_error}
\end{align}
representing the averaged deviation between the \textit{measured} \glspl{tdoa} of the $K-1$ locators and the reference locator $r$, and the \textit{currently predicted} \glspl{tdoa} (\iec \wrt the current estimate $\hat{\mathbf{x}}_{\text{WA}}$) of the $K-1$ locators and the reference locator. With this formulation, the residual error may be interpreted as a measure of how much the estimated distance differences agree with the current position estimate $\hat{\mathbf{x}}_{\text{WA}}$. After computing all residuals, the weights for the next iteration are obtained using Andrews' sine function~\cite{andrews2015robust}
\begin{equation}
w_{T, k} = f_e(e_k) = 
\begin{cases}
 \dfrac{e_\text{max}}{e_k \pi}\cdot \sin\left(\dfrac{e_k \pi}{e_\text{max}}\right) & \text{if } e_k \leq e_{\text{max}} \\[10pt]
0 & \text{else} 
\end{cases} \; 
\label{eq:andrews_weighting}
\end{equation}
with $e_\text{max}$ being the maximum value of the residual error above which the weights are set to 0. In this case, the corresponding locator is discarded. 
After each iteration, a convergence check is performed by comparing
\begin{align}
\Delta = \lVert \mathbf{\hat{x}}_{\text{WA}}^{i} -\mathbf{\hat{x}}_{\text{WA}}^{i-1} \rVert \;
\label{eq:convergence}
\end{align}
to a pre-defined threshold $\epsilon$. In case $\Delta > \epsilon$, we repeat the procedure from the beginning or until a maximum number of iterations $N_\text{it}$ is reached.
Algorithm~\ref{alg:TDoA_IRLS_algo} summarizes the entire \gls{irls} procedure. 
\\As outputs, an initial \gls{tdoa} position estimate $\mathbf{\hat{x}}_{T, \text{IRLS}}$ and locator weights $\mathbf{w}_T$ are provided. Note that, while Algorithm~\ref{alg:TDoA_IRLS_algo} describes a \gls{tdoa} approach, the \gls{tdoa} measurements obtained with \eqref{eq:distance_differences} clearly directly depend on the measured \glspl{toa}. We can therefore use $\mathbf{\hat{x}}_{T, \text{IRLS}}$ and $\mathbf{w}_T$ as initial search point and weights, respectively, for the optimization of the \gls{toa} \gls{ml} function in \eqref{eq:toa_ll_max} and \eqref{eq:joint_ll_fun}.
\\Moreover, note that the complexity is quadratic \wrt the number of measurements, which is reasonable, since typically only few locators are available in a given area. Otherwise, one can choose only a subset of $K^{'}, \; K^{'} \ll K$ locators. The selection should be made based on an a-priori assessment of the reliability of the locators, \egc by choosing the $K^{'}$ ones with the lowest distance or highest \gls{rss} measurements.
\begin{algorithm}[t]
 \caption{\gls{irls} \gls{tdoa} Positioning}\label{alg:TDoA_IRLS_algo}
 \begin{algorithmic}[1]
 \renewcommand{\algorithmicrequire}{\textbf{Input:}}
 \renewcommand{\algorithmicensure}{\textbf{Output:}}
 \REQUIRE estimated distances $\mathbf{\hat{d}} = (\hat{d}_1, \hat {d}_2, \hdots, \hat{d}_K)$, \\
 locator positions $\mathbf{P} = (\mathbf{p}_1, \mathbf{p}_2, \hdots, \mathbf{p}_K)$, \\
  number of iterations $N_{\mathrm{it}}$,
  stopping criterion $\epsilon$, \\
  maximum residual error $e_{\text{max}}$
 \ENSURE \gls{irls} \gls{tdoa} position estimate $\mathbf{\hat{x}}_{T, \text{IRLS}}$,  \\
 locator weights $\mathbf{w}_T$ \\
  \STATE $i \gets 0$
  \STATE get position estimates $\hat{\mathbf{X}}$ with (\ref{eq:TDoA_LES}) - (\ref{eq:ls_position})
  \STATE estimate initial position $\mathbf{\hat{x}}_{\text{WA}}^0$ using (\ref{eq:weighted_position})
  \WHILE {$i \leq N_{\mathrm{it}}$ and $\Delta > \epsilon$}
  \STATE $i \gets i+1$
  \STATE get residual errors $\mathbf{e}$ from (\ref{eq:res_error})
    \STATE update $\mathbf{w}_T$ using (\ref{eq:andrews_weighting})
    \STATE get normalized weights $\bar{\mathbf{w}}_T$ using (\ref{eq:weight_norm})
    \STATE get new estimate $\mathbf{\hat{x}}_{\text{WA}}^i$ according to (\ref{eq:weighted_position})
    \STATE re-compute position estimates $\hat{\mathbf{X}}$  with (\ref{eq:TDoA_LES}) - (\ref{eq:ls_position})  
  \STATE compute $\Delta$ with (\ref{eq:convergence})
  \ENDWHILE
  \STATE $\mathbf{\hat{x}}_{T, \text{IRLS}} \gets \mathbf{\hat{x}}_{\text{WA}}^i$
 \RETURN $\mathbf{\hat{x}}_{T, \text{IRLS}}$, $\mathbf{w}_T$
 \end{algorithmic}
 \end{algorithm}

\subsection{\gls{irls} \gls{aoa} Algorithm}\label{sec:irls_aoa}

The \gls{irls} \gls{aoa} algorithm from \cite{abrudan_2018} is also based on an iterative routine and provides an initial \gls{aoa}-based position estimate $\mathbf{\hat{x}}_{\angle, \text{IRLS}}$ and angular weights $\mathbf{w}_{\angle} = \left(w_{\angle, 1}, w_{\angle, 2}, \ldots\, w_{\angle, K}\right)$ for the $K$ locators. Analogous to the outputs of the \gls{tdoa} algorithm, $\mathbf{\hat{x}}_{\angle, \text{IRLS}}$ and $\mathbf{w}_{\angle}$ are utilized for the optimization of the \gls{aoa} \gls{ml} function in \eqref{eq:vmf_ll_max} and \eqref{eq:joint_ll_fun}.
\\Note that both \gls{irls} algorithms can also be employed as standalone positioning solutions, but in this work we only use them to improve the \gls{ml} techniques introduced in Section \ref{sec:prob_pos} by avoiding local maxima caused by outliers.

\subsection{Joint Positioning Initialization Routine}\label{sec:joint_pos}
For initializing the gradient-based optimization of the joint \gls{toa}/\gls{aoa} \gls{ml} function given in \eqref{eq:joint_ll_fun}, we combine the initial points $\mathbf{\hat{x}}_{T, \text{IRLS}}$ and $\hat{\mathbf{x}}_{\angle, \text{IRLS}}$. We assume the two position estimates to be independent measurements with additive Gaussian noise and diagonal covariance matrices. Therefore, the initial joint location estimate, to be used as the initial search point, can be estimated according to \gls{ml} estimation theory \cite{spagnolini2018statistical} as
\begin{align}
\hat{\mathbf{x}}_{\cap}^{0} 
& = \frac{\hat{\sigma}_{T}^{-2} \cdot \hat{\mathbf{x}}_{T, \text{IRLS}} + \hat{\sigma}_{\angle}^{-2} \cdot \hat{\mathbf{x}}_{\angle, \text{IRLS}}}{\hat{\sigma}_{T}^{-2} + \hat{\sigma}_{\angle}^{-2}}   \; 
\label{eq:joint_pos_init} 
\end{align}
where $\hat{\sigma}_{T}^{2}$ and $\hat{\sigma}_{\angle}^{2}$ are the estimated variances at the respective \gls{tdoa} and \gls{aoa} \gls{irls} position estimates. The variances are obtained from the inverse of the Hessian of the \gls{ll} functions \eqref{eq:toa_ll} and \eqref{eq:vmf_ll}, evaluated at $\mathbf{\hat{x}}_{T, \text{IRLS}}$ and $\hat{\mathbf{x}}_{\angle, \text{IRLS}}$, respectively.
Notice that more elaborate techniques exist to fuse the two position estimates, \egc covariance intersection, which does not make the independence assumption \cite{julier1997non}. However, those algorithms typically come with added complexity, and we found \eqref{eq:joint_pos_init} to be sufficient to obtain a good performance.

We make further use of the variances by discarding the \gls{toa} or \gls{aoa} part of \eqref{eq:joint_ll_fun} if the corresponding variance at the initial position estimates $\hat{\mathbf{x}}_{T, \text{IRLS}}$ and $\hat{\mathbf{x}}_{\angle, \text{IRLS}}$ exceeds a pre-defined threshold $\sigma_{\text{max}}^2$.
For joint \gls{toa}/\gls{aoa} positioning, we now use the final sets of locator weights $\mathbf{w}_{T}$ and $\mathbf{w}_{\angle}$ for time and angle, respectively.  

\section{Experimental Setup}\label{sec:exp_setup}
\begin{figure}[t]
\centering
	\input{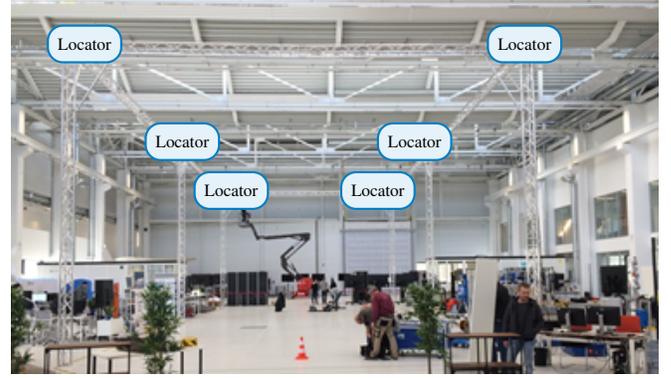}
\caption{Experimental setup with six locators in the industrial research campus ARENA2036. The area under investigation spans roughly 20 x 10 meters and the locators are mounted at a height of about 7 meters.}
\label{fig:ARENA_setup}
\end{figure}

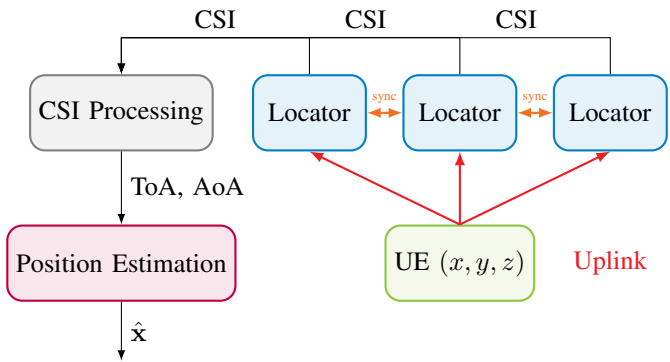
\begin{figure}[h]
  \centering
  \begin{tikzpicture}
\node[fill=mittelblau!10, draw=mittelblau, thick, rounded corners=.2cm,minimum height=1cm,minimum width =1.5cm] (Locator1) at (0.5,0) {Locator};
\node[fill=mittelblau!10, draw=mittelblau, thick, rounded corners=.2cm,minimum height=1cm,minimum width =1.5cm] (Locator2) at (2.5,0) {Locator};
\node[fill=mittelblau!10, draw=mittelblau, thick, rounded corners=.2cm,minimum height=1cm,minimum width =1.5cm] (Locator3) at (4.5,0) {Locator};
  
\node[fill=apfelgruen!10, draw=apfelgruen, thick, rounded corners=.2cm,minimum height=1cm,minimum width =1.5cm] (UE) at (2.5,-2) {UE $(x,y,z)$};

\node[fill=mittelgrau!10, draw=mittelgrau, thick, rounded corners=.2cm,minimum height=1cm,minimum width =1.5cm] (Pros) at (-2,0) {CSI Processing};
\node[fill=purple!10, draw=purple, thick, rounded corners=.2cm,minimum height=1cm,minimum width =1.5cm] (Pos) at (-2,-2) {Position Estimation};

\draw [thick,latex-latex,orange] (Locator1.east)  -- node[above] {\tiny sync} (Locator2.west);
\draw [thick,latex-latex,orange] (Locator2.east)  -- node[above] {\tiny sync} (Locator3.west);

\draw [thick,-latex,rot] (UE.north)  --  (Locator1.south);
\draw [thick,-latex,rot] (UE.north)  --  (Locator2.south);
\draw [thick,-latex,rot] (UE.north)  --  (Locator3.south);
\node[draw=none,rot] (signal) at (4.5,-2) {Uplink};

\draw[-latex] (Locator1.north) --(0.5,1) -- node[above] {CSI}(-2,1) -- (Pros.north);
\draw[-latex] (Locator2.north) -- (2.5,1) -- node[above,xshift=1cm] {CSI}(-2,1) -- (Pros.north);
\draw[-latex] (Locator3.north)  --(4.5,1) -- node[above,xshift=2cm] {CSI}(-2,1) -- (Pros.north);

\draw[-latex] (Pros.south)  -- (Pos.north) node[right, pos=0.5] {\gls{toa}, \gls{aoa}};
\draw[-latex] (Pos.south)  -- (-2,-3.3) node[right, pos=0.5] {$\hat{\mathbf{x}}$};


\end{tikzpicture}
\caption{Block diagram of our \gls{5G} positioning setup.}
  \label{fig:setup}
\end{figure}

\begin{figure}[htb]
\centering
	\input{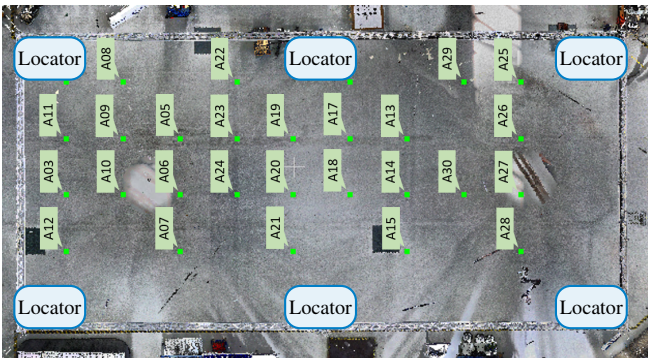}
\caption{Top-down view of the area under investigation including the six locators. The positioning accuracy is evaluated at the 28 ground truth points (small green squares).}
\label{fig:gt_points}
\end{figure}

The performance of our positioning algorithms is evaluated with an experimental lab setup installed at the industrial research campus ARENA2036\footnote{https://www.arena2036.de/en/}. The testbed (Fig. \ref{fig:ARENA_setup}) comprises six tightly synchronized (sub-nanosecond accuracy) locators that are mounted at a height of roughly 7~m and capable of sniffing 5G signals.
Each locator is equipped with a planar antenna array, enabling the estimation of azimuth and elevation \glspl{aoa}. The estimated \gls{csi} provided from each locator is centrally processed by a server that handles the estimation of the three parameters of interest, as well as the computation of the position estimates, according to the schematic in Fig.~\ref{fig:setup}.
The \gls{ue} to be localized transmits a 5G \gls{ofdm} signal at carrier frequency $f_c$ = 3.75~GHz with a bandwidth  $B$ of roughly 100~MHz. The accuracy of our algorithms is evaluated by computing the \gls{2D} positioning errors \wrt the 28 ground truth points depicted in Fig. \ref{fig:gt_points}. 
\section{Results}\label{sec:results}

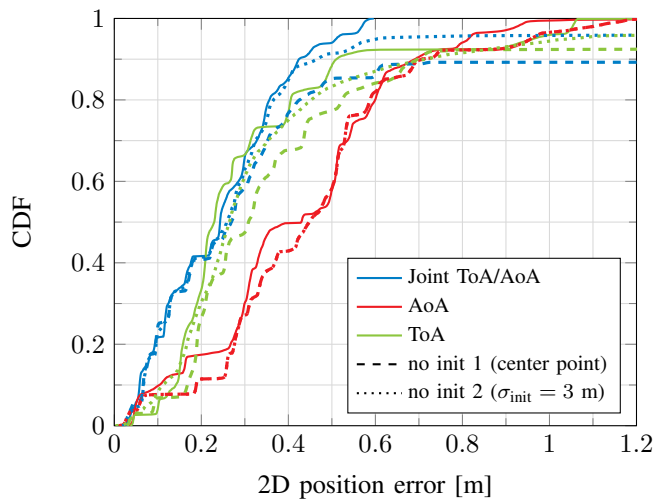
\begin{figure}[t]
  \centering
  \tikzset{mark size=2.5}    
    \begin{tikzpicture}
	\begin{axis}[
    		height = 7cm,
    		width=0.47\textwidth,
    		xlabel={2D position error [m]},
    		ylabel={CDF},
		xmin=0,
		xmax=1.2,
		minor tick num=1,
		yminorticks = true,
		enlargelimits = false,
		legend pos = south east,
		legend style={font=\footnotesize},
		grid = both,
    		grid style={solid, black!15},
		legend cell align={left},
		every axis plot/.append style={very thick},
	    mark repeat={50}
		]

\addplot[thick, color=mittelblau] plot table[each nth point=1, x index=1, y index=0, col sep=comma]{data/joint.txt}; \addlegendentry{Joint \gls{toa}/\gls{aoa}}
\addplot[thick, color=rot] plot table[each nth point=1, x index=1, y index=0, col sep=comma]{data/aoa.txt}; \addlegendentry{\gls{aoa}}
\addplot[thick, color=apfelgruen] plot table[each nth point=1, x index=1, y index=0, col sep=comma]{data/toa.txt}; \addlegendentry{\gls{toa}}

\addplot[thick, color=black, dashed, draw=none] coordinates {(0, 1)}; \addlegendentry{no init 1 (center point)}
\addplot[thick, color=black, dotted, draw=none] coordinates {(0, 1)}; \addlegendentry{no init 2 ($\sigma_{\text{init}} = 3$~m)}

\addplot[very thick, color=mittelblau, dashed] plot table[each nth point=1, x index=1, y index=0, col sep=comma]{data/joint_no_init_center_point.txt};
\addplot[very thick, color=rot, dashed] plot table[each nth point=1, x index=1, y index=0, col sep=comma]{data/aoa_no_init_center_point.txt}; 
\addplot[very thick, color=apfelgruen, dashed] plot table[each nth point=1, x index=1, y index=0, col sep=comma]{data/toa_no_init_center_point.txt};

\addplot[very thick, color=mittelblau, dotted] plot table[each nth point=1, x index=1, y index=0, col sep=comma]{data/joint_no_init_std_dev_3.txt};
\addplot[very thick, color=rot, dotted] plot table[each nth point=1, x index=1, y index=0, col sep=comma]{data/aoa_no_init_std_dev_3.txt}; 
\addplot[very thick, color=apfelgruen, dotted] plot table[each nth point=1, x index=1, y index=0, col sep=comma]{data/toa_no_init_std_dev_3.txt};

\end{axis} 

\end{tikzpicture}
	
		
\caption{CDF of the 2D position error for the three algorithms (\gls{toa}, \gls{aoa}, and joint \gls{toa}/\gls{aoa}). The proposed robust \gls{ml} initialization routine (represented by the solid lines) exhibits no heavy tail. By contrast, the lack of a robust initialization leads to large outliers (dashed and dotted lines).} 
  \label{fig:CDF}
\end{figure}

\begin{figure}[ht]
  \centering
  \begin{tikzpicture}
\tikzstyle{box} = [draw,rounded corners=.1cm,inner sep=5pt,minimum height=4em, text width=5em, align=center, very thick]
\node[box, orange, label={[orange, above]: \footnotesize central area}] (tbd) at (3.7,1.82) {};
	\begin{axis}[
		compat=newest,
		width=0.49\textwidth, height=5.2cm,
		grid=major,
		xlabel={x [m]},legend style={at={(0.5,1.24)},anchor=north},legend columns=2,
		ylabel={y [m]},
        legend entries={Ground Truth,Estimated Position},x dir=reverse,xtick={65,70,75,80},xticklabels={80,75,70,65},
        ylabel shift = -2.5pt,
        xlabel shift = -2.5pt
		]
		\addplot+[black,each nth point=1,only marks,mark=o,scatter,mark size=2pt,scatter/use mapped color={black}] table[x index=1,y index=0]{./data/ground_truth.txt};
		\addplot+[mittelblau,each nth point=1,only marks,mark=+,scatter,mark size=2pt,scatter/use mapped color={mittelblau}] table[x index=1,y index=0,]{./data/joint_pos_est.txt};
		
	\end{axis}
\end{tikzpicture}
\caption{Ground truth (black circles) and estimated (blue crosses) positions using joint \gls{toa}/\gls{aoa} positioning with initialization.}
  \label{fig:area_acc}
\end{figure}
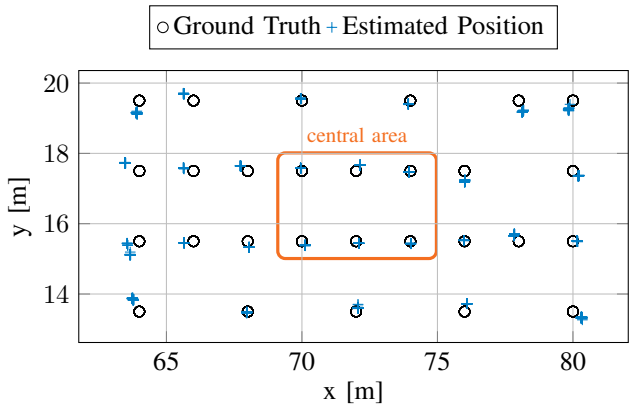

In this section, we compare the investigated algorithms' \gls{2D} positioning accuracy using our previously described experimental setup. Table \ref{tab:params} lists the parameters used for our initialization routine.
\begin{table}[b]
\setlength\extrarowheight{2.5pt}
	\caption{Algorithm Parameters.
	\label{tab:params}}
	\centering
     \begin{tabular}{|c|c|}
       		\hline
       		Maximum iterations of \gls{irls} \gls{tdoa} algorithm $N_\text{{it}}$ & 10 \\
  			\hline
  			Maximum residual error $e_{\text{max}}$ & 2.5 m \\
  			\hline
  			Stopping criterion $\epsilon$ & $10^{-5}$ m \\
  			\hline
		    Maximum variance of position estimate $\sigma_{\text{max}}^2$ & 10 m\textsuperscript{2} \\
      		\hline
  			Maximum concentration parameter $\kappa_{\text{max}}$  & 10 \\
  			\hline
    \end{tabular}
\end{table}
Our proposed robust initialization in conjunction with \gls{ml} positioning is compared to two baselines using the same probabilistic positioning algorithms, but no initialization. In both cases, the available measurements from all locators are always used and equally contribute to the respective \gls{ll}-functions (\iec they have unit weights). The first baseline ``no init 1'' uses the center point of the investigation area as an initial guess, while the second baseline ``no init 2'' assumes knowledge about the \gls{ue}'s rough position. Such information could \egc be obtained via tracking techniques. For ``no init 2'', we model the \gls{2D} uncertainty of the initial point with a standard deviation $\sigma_{\text{init}} = 3$~m, corresponding to the root-mean-square deviation of non-initialized \gls{toa} position estimates.
\\Analyzing the \gls{cdf} curves in Fig. \ref{fig:CDF}, one can observe that joint \gls{toa}/\gls{aoa} \gls{ml} is the most robust technique, clearly outperforming \gls{ml} positioning relying only on \gls{toa} or \gls{aoa} measurements. Moreover, the benefits of our initialization routine become clear, as the baselines without initialization generally perform slightly worse and, more importantly, exhibit heavy tails. On the other hand, our proposed initialization routine allows to achieve a 2D positioning error below 0.6~m in all cases. Similar observations can be made by comparing \gls{toa}-only and \gls{aoa}-only positioning with and without initialization. Also here, initialization slightly improves the overall performance and shows the capability to limit outliers.
\\Table \ref{tab:pos_performance} compares the position error of our proposal and ``no init 2'' \wrt their mean and at different percentiles. Analyzing these values, one can see that the initialization routine has only a small impact on the performance under good conditions, as the median error only slightly improves. However, analyzing higher percentiles, the benefits of proper initialization become clear. While ``no init 2'' exhibits position estimates that can deviate very far from the true position (\egc up to over 10~m for \gls{toa}-only), we are able to keep the \gls{2D} position error below 1.06~m in 99\% of all measurements for all three algorithms when our initialization routine is applied. 
\begin{table}[t]

\definecolor{Gray}{gray}{0.85}
\newcolumntype{g}{>{\columncolor{Gray}}c}

\setlength\extrarowheight{6pt}
	\caption{2D position error summary. \label{tab:pos_performance}}
	\centering
	\setlength\tabcolsep{5pt} 
     \begin{tabular}{p{1.65cm}?c|>{\columncolor[gray]{0.8}}c?c|>{\columncolor[gray]{0.8}}c?c|>{\columncolor[gray]{0.8}}c?}
            &
            \multicolumn{2}{c?}{\textbf{Joint \gls{toa}/\gls{aoa}}} & \multicolumn{2}{c?}{\textbf{\gls{toa}}} &
            \multicolumn{2}{c?}{\textbf{\gls{aoa}}} \\
			\hhline{~|-|-|-|-|-|-|}       		
       		 & init & no init 2 & init & no init 2 & init & no init 2\\
  			\Xhline{2\arrayrulewidth}
  		    \textbf{Mean [m]}& 0.24 & 0.49 & 0.31 & 0.68 & 0.42 & 0.45 \\
  			\hline
  		    \textbf{Median [m]}& 0.24 & 0.26 & 0.23 & 0.26 & 0.43 & 0.45 \\
  			\hline
  		    \textbf{95th \%ile [m]}& 0.51 & 0.60 & 0.96 & 1.05 & 0.81 & 0.96 \\
  			\hline
  		    \textbf{99th \%ile [m]}& 0.57 & 6.35 & 1.06 & 10.24 & 0.94 & 1.15 \\
  			\Xhline{2\arrayrulewidth}

    \end{tabular}
\end{table}
This further highlights the importance of rejecting outliers, as they can cause the position estimate to diverge far from the ground truth, even with a-priori knowledge about the rough position. Remarkably, joint \gls{toa}/\gls{aoa} positioning achieves errors of 0.51~m and 0.57~m at the 95th~\%ile and 99th~\%ile, respectively. The performance at high percentiles is crucial, since the deployment of indoor positioning demands extreme reliability due to requirements for people safety.
\\Lastly, Fig. \ref{fig:area_acc} depicts the distribution of the estimated positions obtained with joint \gls{toa}/\gls{aoa} positioning \wrt the 28 ground truth points.
Analyzing the distribution of the blue crosses, one can infer that our system - even at points where the error is higher - provides stable position estimates without large fluctuations. Moreover, it is clearly visible that the errors at the outer points of the investigation area tend to be higher. This is due to the fact that locators can lie outside of the coverage area at some ground truth points. In these cases, the signal at those locators can be weak and/or arrive at unfavorable (\iec large) \glspl{aoa}. Limiting our analysis only to the six ground truth points of the central area (see orange box in Fig. \ref{fig:area_acc}), we are able to achieve a mean and maximum error of $0.12$~m and $0.23$~m, respectively.



\section{Conclusion}\label{sec:conclusion}
In this work, we presented a robust initialization routine that is applied in an experimental 5G positioning system. The results show that our technique can enhance the performance of probabilistic positioning algorithms by detecting and rejecting outliers and by assigning weights to the measurements based on the confidence in them. For the particular case of joint \gls{toa}/\gls{aoa} \gls{ml} positioning, we are able to avoid large errors at high percentiles, achieving a positioning error of not more than roughly 0.5~m for 95\% of the measurements. Moreover, we showed that joint probabilistic positioning with initialization accomplishes precise localization under good conditions, exhibiting a mean error of 0.12~m for the points located in the central area. 
\\While our findings underline the potential of \gls{5G} indoor positioning, there are still open questions to be addressed. Those include the relationship between locator deployment density and position error, and the system performance in harsh radio environments with more severe multipath propagation.





\section*{Acknowledgments}
This work has been partly funded by the European Commission through the H2020 project Hexa-X (Grant Agreement no. 101015956).

\bibliographystyle{IEEEtran}
\bibliography{5G_Positioning_Main}

\end{document}